# Outer Bounds on the Admissible Source Region for Broadcast Channels with Dependent Sources

Gerhard Kramer
Dept. of Electrical Engineering
University of Southern California
Los Angeles, CA 90089
gkramer@usc.edu

Yingbin Liang
Dept. of Electrical Engineering
University of Hawaii
Honolulu, HI 96822
yingbinl@hawaii.edu

Shlomo Shamai (Shitz)
Dept. of Electrical Engineering
Technion - Israel Institute of Technology
Technion City, Haifa 32000, Israel
sshlomo@ee.technion.ac.il

***Abstract*— Outer bounds on the admissible source region for broadcast channels with dependent sources are developed and used to prove capacity results for several classes of sources and channels.**

## I. Introduction

Consider a two-receiver broadcast channel (BC), say $P_{YZ|X}(\cdot)$, and a discrete memoryless source $(S,T)$ with finite alphabet $\mathcal{S}\times\mathcal{T}$. Source $(S,T)$ is said to be *admissible* for this BC if for any $\lambda$, $0<\lambda<1$, and for large enough $n$ there is a code with length-$n$ codewords such that $P_{e1}\le\lambda$ and $P_{e2}\le\lambda$, where $P_{e1}$ and $P_{e2}$ are the respective error probabilities for receivers 1 and 2. The set of all admissible sources is called the *admissible source region*.

Han and Costa [1] developed a coding strategy that admits certain sources. Let $K=f(S)=g(T)$ be the common variable in the sense of Gacs and Körner (and also Witsenhausen), and consider auxiliary random variables $W,U,V$ that satisfy the Markov chain property

$$ST - WUV - X - YZ. \tag{1}$$

Then the source $(S,T)$ is admissible if (see [1], [2])

$$H(S) < I(SWU;Y) - I(T;WU|S) \tag{2}$$
$$H(T) < I(TWV;Z) - I(S;WV|T) \tag{3}$$
$$H(ST) < \min\{I(KW;Y),I(KW;Z)\} + I(SU;Y|KW) + I(TV;Z|KW) - I(SU;TV|KW) \tag{4}$$
$$H(ST) < I(SWU;Y) + I(TWV;Z) - I(SU;TV|KW) - I(ST;KW). \tag{5}$$

Our paper is concerned with ”outer bounds” on the set of admissible sources, i.e., we wish to determine a superset of the admissible source region. To do so, we borrow tools from [3] and modify them to include dependent sources. We remark that a more general class of outer bounds for BCs is presented in [4]. Nevertheless, we use the approach of [3] because the auxiliary random variables are simpler and they can be used to prove capacity theorems. For example, it is not obvious how to choose the auxiliary random variables of [4] to even determine the capacity region of degraded broadcast channels.

## II. Outer Bounds on the Admissible Source Region

Consider the source sequences $S^m,T^m$ where $m$ may be different than the code length $n$. Let $R=n/m$ be the source-channel “bandwidth expansion”. Further let the auxiliary random variables $(\widetilde{K},\widetilde{S},\widetilde{T})$ have the same distribution as $(K^m,S^m,T^m)$ and let them form the Markov chain

$$\widetilde{K}\widetilde{S}\widetilde{T}UV - X - YZ. \tag{6}$$

We may also let $X$ be a deterministic function of $(\widetilde{S},\widetilde{T})$. We have the following outer bound.

*Theorem 1:* An admissible source $(S,T)$ satisfies the following bounds for some Markov chain (6):

$$H(K)/R \le \min\{I(\widetilde{K};Y|U), I(\widetilde{K};Z|V)\} \tag{7}$$
$$H(S)/R \le I(\widetilde{S};Y|U) \tag{8}$$
$$H(T)/R \le I(\widetilde{T};Z|V) \tag{9}$$
$$H(ST)/R \le I(\widetilde{S};Y|\widetilde{T}UV) + I(\widetilde{T}U;Z|V) \tag{10}$$
$$H(ST)/R \le I(\widetilde{T};Z|\widetilde{S}UV) + I(\widetilde{S}V;Y|U) \tag{11}$$
$$H(ST)/R \le I(\widetilde{K}UV;Y) + I(\widetilde{S};Y|\widetilde{T}UV) + I(\widetilde{T};Z|\widetilde{K}UV) \tag{12}$$
$$H(ST)/R \le I(\widetilde{K}UV;Z) + I(\widetilde{T};Z|\widetilde{S}UV) + I(\widetilde{S};Y|\widetilde{K}UV). \tag{13}$$

*Remark 1:* Consider the case $R=1$, $K,S,T$ statistically independent, $KST$ statistically independent of $UV$, and $(H(K),H(S),H(T))=(R_0,R_1,R_2)$. The bounds (7)-(13) are then a subset of those given in [3, Thm. 1].

*Theorem 2:* A weaker set of conditions than (7)-(13) is as follows, where $Q=(\widetilde{K},U,V)$:

$$H(K)/R \le \min\{I(Q;Y), I(Q;Z)\} \tag{14}$$
$$H(S)/R \le I(Q\widetilde{S};Y) \tag{15}$$
$$H(T)/R \le I(Q\widetilde{T};Z) \tag{16}$$
$$H(ST)/R \le I(\widetilde{S};Y|Q\widetilde{T}) + I(Q\widetilde{T};Z) \tag{17}$$
$$H(ST)/R \le I(\widetilde{T};Z|Q\widetilde{S}) + I(Q\widetilde{S};Y) \tag{18}$$
$$H(ST)/R \le I(Q;Y) + I(\widetilde{S};Y|Q\widetilde{T}) + I(\widetilde{T};Z|Q) \tag{19}$$
$$H(ST)/R \le I(Q;Z) + I(\widetilde{T};Z|Q\widetilde{S}) + I(\widetilde{S};Y|Q). \tag{20}$$

*Remark 2:* Consider the case described in Remark 1. The bounds (14)-(20) are then identical to those in [3, Thm. 3].

## III. Examples

We continue by considering only $R = 1$ for simplicity.

### A. Separating Source and Channel Coding

A natural approach is to decouple $ST$ from $WUV$, i.e., choose $WUV$ independent of $ST$. Effectively, we convert the source strings to (compressed) bit strings and apply channel coding. The bounds (2)-(5) are then

$$H(S) < I(WU;Y) \tag{21}$$

$$H(T) < I(WV;Z) \tag{22}$$

$$H(ST) < \min\{I(W;Y), I(W;Z)\} + I(U;Y|W) + I(V;Z|W) - I(U;V|W) - I(S;T|K) \tag{23}$$

$$H(ST) < I(WU;Y) + I(WV;Z) - I(U;V|W) - I(S;T|K) - I(ST;K). \tag{24}$$

### B. Markov Sources

Consider the case where $S - K - T$ forms a Markov chain, that is $I(S;T|K) = 0$. This situation seems to be effectively the classic case where there are three independent messages $W_0, W_1, W_2$ with $nR_0, nR_1, nR_2$ bits, respectively. In fact, for any BC for which we know the capacity region, we can match the outer bound of Theorem 2 to the admissible bounds (21)-(24). For example, consider a semi-deterministic BC where $Y$ is a deterministic function of $X$. We have the following result.

*Theorem 3:* A source for which $S - K - T$ forms a Markov chain is admissible for a semi-deterministic BC where $Y$ is a deterministic function of $X$ if $ST$ satisfies

$$H(K) < \min\{I(W;Y), I(W;Z)\} \tag{25}$$

$$H(S) < H(Y) \tag{26}$$

$$H(T) < I(WV;Z) \tag{27}$$

$$H(ST) < H(Y|WV) + I(VW;Z) \tag{28}$$

$$H(ST) < I(W;Y) + H(Y|WV) + I(V;Z|W) \tag{29}$$

for some Markov chain $ST - WV - X - YZ$. Conversely, such a source is not admissible for such a BC if for every Markov chain $ST - WV - X - YZ$ the direction of one of the (strict) inequalities is reversed.

*Proof:* For admissibility, choose $U = Y$ and observe that the bounds (25)-(29) imply the bounds (21)-(24). For the converse, consider (14)-(17) and (19), identify $Q, \widetilde{K}, \widetilde{T}$ with $W, K, V$, respectively, and use

$$I(Q\widetilde{S};Y) \leq H(Y) \tag{30}$$

$$I(\widetilde{S};Y|Q\widetilde{T}) \leq H(Y|Q\widetilde{T}). \tag{31}$$

■

### C. More Capable BCs

Theorem 3 does not give the admissible source region since it requires $S - K - T$ to form a Markov chain. In fact, Han and Costa showed in [1, Example 2] that separating source and channel coding (implied by decoupling $ST$ and $WUV$) is suboptimal in general. The decoupling also doesn't *seem* to work for the expressions (2)-(5) for an important class of channels where one might guess that it should.

Consider the class of more-capable BCs defined by the constraint that

$$I(X;Y) \geq I(X;Z) \text{ for all } P_X. \tag{32}$$

This class includes both physically and stochastically degraded BCs. If we choose $WUV$ independent of $ST$ then (21)-(24) still exhibit the rate loss $I(S;T|K)$ in (23)-(24). However, choosing $W$ differently we have the following result.

*Theorem 4:* The admissible source region for more-capable BCs is defined by the bounds

$$H(ST) \leq I(X;Y) \tag{33}$$

$$H(T) \leq I(\widetilde{W};Z) \tag{34}$$

$$H(ST) \leq I(X;Y|\widetilde{W}) + I(\widetilde{W};Z) \tag{35}$$

for some Markov chain $ST - \widetilde{W} - X - YZ$.

*Proof:* We choose

$$W = T\widetilde{W} \text{ where } ST \text{ is independent of } \widetilde{W}UV \tag{36}$$

$$U = X = \text{a noisy function of } \widetilde{W} \tag{37}$$

$$V = \text{a constant.} \tag{38}$$

The bounds (2)-(5) are then simply (33)-(35). Furthermore, for any Markov chain $ST - \widetilde{W} - X - YZ$ we can achieve the right-hand sides of (33)-(35) since these depend on $ST$ only through $\widetilde{W}$. For the converse, consider (16)-(18) and identify $Q\widetilde{T}, \widetilde{S}$ with $\widetilde{W}, U$, respectively. The bounds (33)-(35) follow from (16)-(18) and the bounds

$$\begin{aligned} I(\widetilde{T};Z|Q\widetilde{S}) &\leq I(X;Z|Q\widetilde{S}) \\ &= \sum P_{Q\widetilde{S}}(ab) I(X;Z|Q\widetilde{S} = ab) \\ &\leq \sum P_{Q\widetilde{S}}(ab) I(X;Y|Q\widetilde{S} = ab) \\ &= I(X;Y|Q\widetilde{S}) \end{aligned} \tag{39}$$

where the second inequality follows by applying (32). ■

We remark that the choice $W = T\widetilde{W}$ does, in fact, permit separating source and channel coding because $\widetilde{W}$ is independent of $T$. The coding approach is to simply compress $T$ to its entropy-rate $H(T)$ and consider the resulting bits as a common message. Next, compress $S$ to the conditional entropy-rate $H(S|T)$ and consider the resulting bits as a private message. Decoder $Y$ first decodes and decompresses $T$ and then decodes and decompresses $S$. Decoder $Z$ decodes and decompresses $T$ only. This natural approach is included in the Han-Costa coding method, but only indirectly.

## IV. Proof of Theorem 1

Let $|\mathcal{S}|$ be the cardinality of the set $\mathcal{S}$ and let $(\widetilde{K}, \widetilde{S}, \widetilde{T}) = (K^m, S^m, T^m)$. Fano's inequality [5, Sec. 2.11] gives

$$H(K^m|Y^n) \leq H(S^m|Y^n) \leq P_{e1} \cdot m \log_2 |\mathcal{S}| + 1 \tag{40}$$

$$H(K^m|Z^n) \leq H(T^m|Z^n) \leq P_{e2} \cdot m \log_2 |\mathcal{T}| + 1 \tag{41}$$

Let $\delta_1 = P_{e1}\log_2|\mathcal{S}|+1/m$ and $\delta_2 = P_{e2}\log_2|\mathcal{S}|+1/m$, and observe that reliable communication ($P_{e1}\to 0$ and $P_{e2}\to 0$) means that $\delta_1 \to 0$ and $\delta_2 \to 0$ as $m\to\infty$. We define the following auxiliary random variables

$$U_i = Y^{i-1}, \quad V_i = Z_{i+1}^n \tag{42}$$

$$P_I(i) = \frac{1}{n},\ i=1,2,\ldots,n \tag{43}$$

$$U = (U_I, I), \quad V = (V_I, I) \tag{44}$$

$$X = X_I, \quad Y = Y_I, \quad Z_I = Z \tag{45}$$

and observe that we have the Markov chain

$$\widetilde{K}\widetilde{S}\widetilde{T}UV - X - YZ. \tag{46}$$

Furthermore, the definition (45) of $XYZ$ is consistent with our BC in the sense that $P_{YZ|X}(\cdot)$ is the same in Sec. I. Fano's inequality (40) implies

$$\begin{aligned} m[H(K)-\delta_1] &\le H(K^m) - H(K^m|Y^n) && (47)\\ &= I(K^m;Y^n) && (48)\\ &= \sum_{i=1}^n I(K^m;Y_i|Y^{i-1}) && (49)\\ &= \sum_{i=1}^n I(\widetilde{K};Y_i|U_i) && (50)\\ &= nI(\widetilde{K};Y_I|U_I I) && (51)\\ &= nI(\widetilde{K};Y|U) && (52) \end{aligned}$$

We similarly have

$$m[H(K)-\delta_2] \le nI(\widetilde{K};Z|V) \tag{53}$$

$$m[H(S)-\delta_1] \le nI(\widetilde{S};Y|U) \tag{54}$$

$$m[H(T)-\delta_2] \le nI(\widetilde{T};Z|V). \tag{55}$$

To develop our other bounds, we will use the following identities (see [6, p. 332] and [7, Lemma 7]).

*Lemma 1:* For any random variables $W, Y^n, Z^n$ we have

$$I(W;Z^n) = \sum_{i=1}^n \Big[I(WY^{i-1};Z_i^n) - I(WY^i;Z_{i+1}^n)\Big] \tag{56}$$

where $Y^0 = Z_{n+1}^n = 0$.

*Proof:* By direct calculation. ■

*Lemma 2:* For any random variables $W, Y^n, Z^n$ we have

$$\sum_{i=1}^n I(Z_i;Y^{i-1}|WZ_{i+1}^n) = \sum_{i=1}^n I(Y_i;Z_{i+1}^n|WY^{i-1}). \tag{57}$$

*Proof:* See [7, Lemma 7]. ■

Consider the following steps:

$$\begin{aligned} &m[H(S)+H(T)-\delta_1-\delta_2] && (58)\\ &\le I(S^m;Y^n) + I(T^m;Z^n) && (59)\\ &\le I(S^m;T^m) + I(S^m;Y^n|T^m) + I(T^m;Z^n) && (60)\\ &= mI(S;T) + \sum_{i=1}^n \Big[I(S^m;Y_i|T^mY^{i-1}) \\ &\qquad + I(T^mY^{i-1};Z_i^n) - I(T^mY^i;Z_{i+1}^n)\Big] && (61) \end{aligned}$$

where the last step follows from Lemma 1. Continuing, we have

$$\begin{aligned} &m[H(ST)-\delta_1-\delta_2] && (62)\\ &= \sum_{i=1}^n \Big[I(S^m;Y_i|T^mY^{i-1}) + I(T^mY^{i-1};Z_i|Z_{i+1}^n) \\ &\qquad - I(Y_i;Z_{i+1}^n|T^mY^{i-1})\Big] && (63)\\ &= \sum_{i=1}^n \Big[-H(Y_i|T^mS^mY^{i-1}) + I(T^mY^{i-1};Z_i|Z_{i+1}^n) \\ &\qquad + H(Y_i|T^mY^{i-1}Z_{i+1}^n)\Big] && (64)\\ &\le \sum_{i=1}^n \Big[-H(Y_i|T^mS^mY^{i-1}Z_{i+1}^n) \\ &\qquad + I(T^mY^{i-1};Z_i|Z_{i+1}^n) + H(Y_i|T^mY^{i-1}Z_{i+1}^n)\Big] && (65)\\ &= \sum_{i=1}^n \Big[I(\widetilde{S};Y_i|\widetilde{T}U_iV_i) + I(\widetilde{T}U_i;Z_i|V_i)\Big] && (66)\\ &= n\Big[I(\widetilde{S};Y|\widetilde{T}UV) + I(\widetilde{T}U;Z|V)\Big]. && (67) \end{aligned}$$

Similar arguments give

$$m[H(ST)-\delta_1-\delta_2] \le n\Big[I(\widetilde{T};Z|\widetilde{S}UV) + I(\widetilde{S}V;Y|U)\Big]. \tag{68}$$

Next, consider the following steps:

$$\begin{aligned} &m[H(S)+H(T)-\delta_1-\delta_2] && (69)\\ &\le I(S^m;Y^n) + I(T^m;Z^n) && (70)\\ &\le I(K^m;Y^n) + I(S^m;Y^n|K^m) \\ &\qquad + I(K^m;T^m) + I(T^m;Z^n|K^m) && (71)\\ &\le I(K^m;Y^n) + I(S^m;T^m|K^m) + I(S^m;Y^n|K^mT^m) \\ &\qquad + I(K^m;T^m) + I(T^m;Z^n|K^m) \\ &\qquad \pm I(T^m;Y^n|K^m) && (72)\\ &= I(K^mS^mT^m;Y^n) - I(T^m;Y^n|K^m) + I(T^m;Z^n|K^m) \\ &\qquad + m(I(S;T|K)+H(K)). && (73) \end{aligned}$$

The reader can check that

$$I(S;T|K) + H(K) = I(S;T). \tag{74}$$

The first expression in (73) is bounded as

$$\begin{aligned} I(K^mS^mT^m;Y^n) &= \sum_{i=1}^n I(K^mS^mT^m;Y_i|Y^{i-1}) && (75)\\ &= \sum_{i=1}^n I(\widetilde{K}\widetilde{S}\widetilde{T};Y_i|U_i) && (76)\\ &\le \sum_{i=1}^n I(\widetilde{K}\widetilde{S}\widetilde{T}U_iV_i;Y_i) && (77) \end{aligned}$$

The second and third expressions in (73) can be manipulated as follows:

$$- I(T^m; Y^n|K^m) + I(T^m; Z^n|K^m) \tag{78}$$

$$= \sum_{i=1}^{n} \Big[ - I(T^m; Y_i|K^m Y^{i-1}) + I(T^m; Z_i|K^m Z_{i+1}^n) \Big] \tag{79}$$

$$= \sum_{i=1}^{n} \Big[ - I(T^m Z_{i+1}^n; Y_i|K^m Y^{i-1}) + I(Z_{i+1}^n; Y_i|K^m T^m Y^{i-1}) + I(T^m Y^{i-1}; Z_i|K^m Z_{i+1}^n) - I(Y^{i-1}; Z_i|K^m T^m Z_{i+1}^n) \Big] \tag{80}$$

$$= \sum_{i=1}^{n} \Big[ - I(T^m Z_{i+1}^n; Y_i|K^m Y^{i-1}) + I(T^m Y^{i-1}; Z_i|K^m, Z_{i+1}^n) \Big] \tag{81}$$

where the last step follows from Lemma 2. Continuing, we have

$$- I(T^m; Y^n|K^m) + I(T^m; Z^n|K^m) \tag{82}$$

$$= \sum_{i=1}^{n} \Big[ - I(Z_{i+1}^n; Y_i|K^m Y^{i-1}) - I(T^m; Y_i|K^m Y^{i-1}, Z_{i+1}^n) + I(Y^{i-1}; Z_i|K^m Z_{i+1}^n) + I(T^m; Z_i|K^m Z_{i+1}^n Y^{i-1}) \Big] \tag{83}$$

$$= \sum_{i=1}^{n} \Big[ - I(T^m; Y_i|K^m Y^{i-1} Z_{i+1}^n) + I(T^m; Z_i|K^m Z_{i+1}^n Y^{i-1}) \Big] \tag{84}$$

$$= \sum_{i=1}^{n} \Big[ - I(T^m; Y_i|K^m U_i V_i) + I(T^m; Z_i|K^m U_i V_i) \Big] \tag{85}$$

where the second last step follows from Lemma 2.

We substitute (74), (77), and (85) into (73) and obtain

$$m[H(ST) - \delta_1 - \delta_2] \tag{86}$$

$$\leq \sum_{i=1}^{n} \Big[ I(\widetilde{K}\widetilde{S}\widetilde{T} U_i V_i; Y_i) - I(\widetilde{T}; Y_i|\widetilde{K} U_i V_i) + I(\widetilde{T}; Z_i|\widetilde{K} U_i V_i) \Big] \tag{87}$$

$$= \sum_{i=1}^{n} \Big[ I(\widetilde{K} U_i V_i; Y_i) + I(\widetilde{S}; Y_i|\widetilde{K}\widetilde{T} U_i V_i) + I(\widetilde{T}; Z_i|\widetilde{K} U_i V_i) \Big] \tag{88}$$

$$= n \Big[ I(\widetilde{K} U_I V_I; Y_I|I) + I(\widetilde{S}; Y_I|\widetilde{T} U_I V_I I) + I(\widetilde{T}; Z_I|\widetilde{K} U_I V_I I) \Big] \tag{89}$$

$$\leq n \Big[ I(\widetilde{K} U V; Y) + I(\widetilde{S}; Y|\widetilde{T} U V) + I(\widetilde{T}; Z|\widetilde{K} U V) \Big] \tag{90}$$

By similar arguments, we also have

$$m[H(ST) - \delta_1 - \delta_2] \leq n \Big[ I(\widetilde{K} U V; Z) + I(\widetilde{T}; Z|\widetilde{S} U V) + I(\widetilde{S}; Y|\widetilde{K} U V) \Big] \tag{91}$$